# Efficient geometrical control of spin waves in microscopic YIG waveguides

S. R. Lake[1], B. Divinskiy[2*], G. Schmidt[1,3], S. O. Demokritov[2], and V. E. Demidov[2]

[1]*Institut für Physik, Martin-Luther-Universität Halle-Wittenberg, 06120 Halle, Germany*

[2]*Institute for Applied Physics, University of Muenster, 48149 Muenster, Germany*

[3]*Interdisziplinäres Zentrum für Materialwissenschaften, Martin-Luther-Universität Halle-Wittenberg, 06120 Halle, Germany*

We study experimentally and by micromagnetic simulations the propagation of spin waves in 100-nm thick YIG waveguides, where the width linearly decreases from 2 to 0.5 μm over a transition region with varying length between 2.5 and 10 μm. We show that this geometry results in a down conversion of the wavelength, enabling efficient generation of waves with wavelengths down to 350 nm. We also find that this geometry leads to a modification of the group velocity, allowing for almost-dispersionless propagation of spin-wave pulses. Moreover, we demonstrate that the influence of energy concentration outweighs that of damping in these YIG waveguides, resulting in an overall increase of the spin-wave intensity during propagation in the transition region. These findings can be utilized to improve the efficiency and functionality of magnonic devices which use spin waves as an information carrier.

*Corresponding author, e-mail: b_divi01@uni-muenster.de

Spin waves propagating in microscopic magnetic waveguides present a flexible and highly functional tool for transmission and processing of information on the nano-scale[1-4]. Among the most important advantages provided by spin waves is their controllability by the magnetic field, which, for example, enables efficient control of their propagation characteristics by electric currents. This controllability also forms the basis of using spatially non-uniform, dipolar magnetic fields to manipulate spin waves[5-7]. Because these fields are determined by the waveguide's geometry, varying its spatial parameters enables different mode transformations and wavelength conversion[7-15]. Although tuning spin waves by using geometrical effects provides many opportunities for the implementation of magnonic devices, the functionality of this approach is strongly limited by the spatial attenuation. Indeed, in metallic waveguides with a small decay length, the passage through a conversion region can lead to a massive loss of spin-wave intensity[16].

The restrictions imposed by the fast spatial decay of spin waves can be overcome by using high-quality, nanometers-thick films of the low-damping magnetic insulator, yttrium iron garnet (YIG)[17-19], where the decay length of spin waves can surpass many tens of micrometers[20-22]. Recently it was shown that these films can be structured on the micrometer and the sub-micrometer scale without significantly increasing the magnetic damping[23-25]. Additionally, magnetic dynamics in these films can be controlled by spin-torque effects, which can be used to enhance further the propagation characteristics[26,27] and generate propagating spin waves by dc electric currents without the need to use energy-inefficient microwave excitation[28]. These features make ultrathin YIG films an excellent candidate for magnonic applications where spin waves are steered via geometrical parameters.

In this Letter, we study the control of spin-wave propagation characteristics in microscopic, ultrathin-YIG waveguides in which the width linearly decreases along the propagation direction. By using spatially-, temporally-, and phase-resolved measurements and micromagnetic simulations, we show that the spatial variation of the demagnetizing field caused by the narrowing of the waveguide results in a robust decrease of the spin-wave wavelength. Due to the minimal spatial attenuation, this wavelength conversion occurs without the decrease of the spin-wave intensity during propagation in the transition region. On the contrary, due to spatial compression, the intensity exhibits a noticeable increase, which becomes particularly pronounced for shorter transition lengths. These effects can be utilized to implement highly efficient excitation of short-wavelength spin waves. Additionally, we show that the geometrical control can be used to tune the propagation velocity of spin-wave pulses and reach a regime where the velocity is almost independent of the spin-wave frequency. Our findings demonstrate a simple and robust method to control spin-wave propagation which can enhance the functionality of nanoscale magnonic devices.

Figure 1(a) shows the schematics of our experiment. We study a microscopic spin-wave waveguide patterned from a 100-nm thick YIG film grown by pulsed-laser deposition (PLD). The YIG film is characterized by a saturation magnetization of $4\pi M = 1.75$ kG and a Gilbert damping constant $\alpha = 4\times10^{-4}$, as determined from ferromagnetic-resonance measurements. The width of the waveguide, $w$, linearly decreases from 2 μm to 0.5 μm over a 10- μm long transition region. The spin waves are excited by using a 500-nm wide and 150-nm thick inductive Au antenna perpendicular to the waveguide, with its right-hand edge located at the beginning of the transition region.

The structures were patterned on a GGG <111> substrate using a two-layer PMMA resist and subsequent electron beam lithography. After development in isopropanol, 110 nm of YIG was deposited via PLD, following a recipe published by Hauser *et al*. (Ref. 19). The sample was then placed in acetone for lift-off of extraneous material and afterwards annealed in a pure oxygen atmosphere[19]. Next, 10 nm of YIG were etched using phosphoric acid in order to remove seams that can appear at the edges of the structures due to the mobility of the deposited atoms during PLD. Finally, the overlying antenna was patterned using a tri-layer PMMA resist, evaporation of Ti (10 nm) and Au (150 nm), and lift-off.

The YIG waveguide is magnetized to saturation by an in-plane, static magnetic field, $H_0$, applied along the Au antenna. Because of demagnetization effects, the internal magnetic field $H_{int}$ is smaller than $H_0$. It is not uniform across the waveguide width and strongly differs in the waveguide's widest and narrowest parts (see the distribution in Fig. 1(b) calculated by using the micromagnetic simulation software MuMax3 (Ref. 29)). At $H_0 = 1000$ Oe, the maximum internal field is 945 and 785 Oe in the widest and the narrowest part, respectively. As seen from Fig. 1(c), this difference results in a shift of approximately 0.7 to 0.8 GHz in the dispersion curves.

We note that the dispersion curves calculated by using MuMax3 (curves in Fig. 1(c)) coincide well with those obtained from phase-resolved measurements (symbols in Fig. 1(c)) described in detail below. This good agreement allows us to rely on results of simulations to obtain the information about the propagation of spin waves which cannot be obtained from direct measurements.

To analyze the propagation of spin waves experimentally, we use the time- and phase-resolved micro-focus Brillouin light scattering (BLS) spectroscopy[16]. We focus the probing laser light with the wavelength of 473 nm and a power of 0.25 mW into a diffraction-limited spot on the surface of the YIG waveguide (see Fig. 1(a)) and analyze the light inelastically scattered from spin waves. The measured signal, or BLS intensity, is proportional to the intensity of spin waves at the position of the probing spot, which allows us to record two-dimensional spin-wave intensity maps. Additionally, by using the interference of the scattered light with the probing light modulated by the microwave excitation signal, we measure the spatial maps of $\cos(\varphi)$, where $\varphi$ is the phase difference between the spin wave and the signal applied to the antenna. The Fourier analysis of these maps provides direct information about the wavelength of spin waves at a given excitation frequency.

Figures 2(a) and 2(b) show representative phase and intensity maps recorded at the excitation frequency $f=4.5$ GHz. The left edge of the maps corresponds to the position $x=0.5$ µm which is selected to avoid measuring where the probing light is partially blocked by the antenna. The data of Fig. 2(a) indicate that the wavelength of spin waves gradually decreases during propagation in the transition region, reflecting the frequency shift in the dispersion spectrum due to the continuous reduction of the internal static magnetic field as seen in Figs. 1(b) and 1(c). We note that the phase profiles are slightly disturbed in the vicinity of the antenna, which is caused by the weak excitation of higher-order transverse waveguide modes[16]. The slight periodic transverse modulation of the intensity distribution seen in Fig. 2(b) also demonstrates this effect.

Analysis of the experimental maps shows that in the transition region the wavelength of the spin wave decreases from about 4 µm to 0.5 µm, i.e., by a factor of 8, (point-down triangles in Fig. 2(c)). This is in quantitative agreement with the results obtained from micromagnetic simulations (point-up triangles in Fig. 2(c)). The wavelength-conversion process is characterized thoroughly in Fig. 2(d), which shows the spin-wave wavelength at the end of the transition region, $\lambda_{OUT}$, as a function of the wavelength of the spin wave excited by the antenna, $\lambda_{EXC}$. We note that the efficiency of the inductive excitation by the 500-nm wide antenna quickly decreases for waves with wavelengths smaller than 1 µm (Ref. 16), limiting the interval of $\lambda_{EXC}$ accessible in the experiment. This restriction is a significant drawback of the inductive excitation mechanism, which strongly limits its use in magnonic devices operating with short-wavelength spin waves. As seen from the data of Fig. 2(d), the observed conversion of the wavelength allows one to extend the range of usable wavelengths down to 350 nm.

From the point of view of technical applications, the demonstrated wavelength conversion is advantageous only if it is not accompanied by a strong decrease of the intensity of the spin wave in the transition region. On one hand, one expects a spatial decrease of the intensity due to the damping and/or wave reflections. On the other hand, the narrowing of the waveguide is expected to result in an increase of the wave's intensity due to its energy being concentrated into a smaller cross section. To prove which of these mechanisms dominates in the studied waveguide, we analyze spatial dependencies of the spin-wave intensity obtained from the measurements and micromagnetic simulations (Fig. 3(a)). The experimental curve in Fig. 3(a) exhibits an almost constant intensity in the interval $x$=0.5-7 μm. This indicates that the energy-concentration effect approximately compensates the effects of the damping. However, at larger $x$, the intensity quickly decreases. We emphasize that this observation cannot be related to wave reflections because the intensity profile shows no signature of the formation of a standing wave. We also note that the observed decrease is not reproduced in the micromagnetic simulations. The calculated intensity coincides well with the experimental one in the range $x$=0.5-7 μm. However, contrary to experimental data, the calculated intensity noticeably increases in the vicinity of the end of the transition region. We associate this discrepancy with the wavelength-dependent sensitivity of the measurement apparatus. Indeed, the sensitivity of magneto-optical techniques is known to decrease with decreasing wavelength of spin waves, $\lambda$, and vanish when the latter becomes equal to the diameter of the probing light spot, $d$. Assuming $d$=0.3 μm and $\lambda$ decreasing from 4 to 0.5 μm, one can estimate that the experimental sensitivity decreases approximately by a factor of 4. This is in good agreement with the ratio between the calculated and experimental intensities at $x>10$ µm in Fig. 3(a). Taking this into account, we base our further analysis on the results of simulations.

Considering the calculated intensity curve (Fig. 3(a)) and comparing the intensity of spin waves at the beginning of the transition region with the intensity at a point 0.5 µm beyond the end of this region, we conclude that the conversion process is accompanied by an increase of the spin-wave intensity by approximately a factor of 1.5. This result clearly proves the suitability of the proposed conversion mechanism for practical applications. Additionally, as shown by the data of Fig. 3(b), the intensity enhancement can be further improved by reducing the length of the transition region. Note here, that this reduction also leads to stronger reflections of the wave at the end of the transition, as seen from the increasing intensity drop at that point (marked by arrows in Fig. 3(b)). However, this adverse effect does not compromise the overall increase of the intensity enhancement (see the inset in Fig. 3(b)).

Finally, we analyze the effects of the spatial variation of the waveguide geometry on the propagation of spin-wave pulses. As can be seen in Fig. 1(c), the reduction of the waveguide width affects not only the wavelength of a spin wave at a given frequency, but also the slope of the dispersion curve, which determines the group velocity. In other words, during the propagation in the transition region, the dispersion of a spin-wave pulse is expected to change as well. In order to address this phenomenon, we perform time-resolved BLS measurements using an excitation signal in the form of 20-ns long pulses and determine the temporal delay of the spin-wave pulses at different spatial positions. In agreement with the above arguments, the found dependence of the propagation delay (Fig. 4(a)) is not linear within the transition region and clearly exhibits a gradual deceleration of the pulse. The observed deceleration can be used, for example, to implement controllable compression of the spin-wave pulses in the space domain.

From the local slope of the dependence shown in Fig. 4(a), we determine the spin-wave group velocities at the beginning of the transition region ($x$=0.5 µm) and at its end ($x$=10 µm). Figure 4(b) summarizes these results obtained at different excitation frequencies. These data show that the initial-stage group velocity depends strongly on the carrier frequency, $f$, while the velocity at $x$=10 µm is almost independent of frequency. The latter observation indicates that the spin-wave pulses experience nearly dispersionless propagation in the narrow ($w$=0.5 µm) waveguide[30]. In narrow waveguides, the spectral region where the group velocity has weak frequency dependence corresponds to spin waves with relatively short wavelengths[30]. These are difficult to excite by an inductive antenna but, by the down conversion presented here, can be easily achieved. The data of Fig. 4(b) show that the demonstrated approach makes it possible to achieve propagation of spin-wave pulses over longer distances without pulse broadening by dispersion effects.

In conclusion, we show that the geometrical control in ultrathin-YIG waveguides provides practical opportunities for spin-wave manipulations, such as the down conversion of the wavelength and the tuning of the propagation velocity. We demonstrate that the intensity of spin waves can be maintained while passing through the control region, due to the small damping in YIG, and can in fact noticeably increase due to spatial compression. These findings can be used for implementation of energy-efficient magnonic devices that exploits sub-micrometer wavelengths.

This work was supported in part by the Deutsche Forschungsgemeinschaft (DFG, German Research Foundation) – Project-ID 433682494 – SFB 1459 and TRR227 TP B02.

## Data availability

The data that support the findings of this study are available from the corresponding author upon reasonable request.

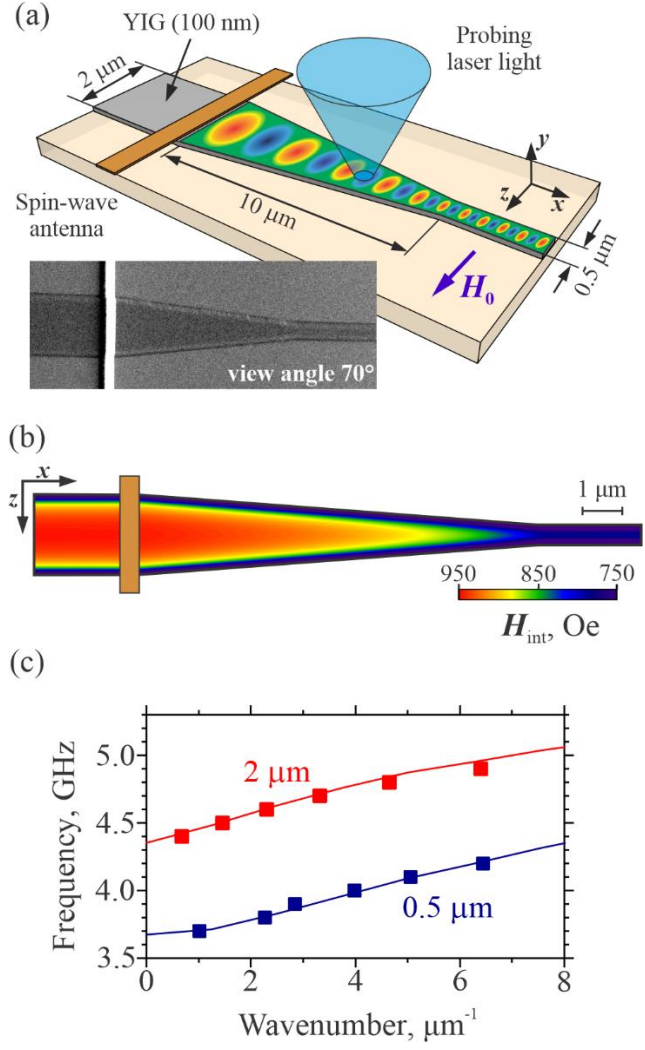

FIG. 1. (a) Schematics of the experiment. Inset shows the scanning-electron micrograph of the sample recorded under an angle of 70°. (b) Distribution of the internal static magnetic field in the waveguide calculated by using micromagnetic simulations. (c) Dispersion curves of spin waves in the widest ($w=2$ μm) and the narrowest ($w=0.5$ μm) parts of the waveguide obtained from micromagnetic simulations (curves) and from phase-resolved measurements (symbols). The data are obtained at $H_0=1000$ Oe.

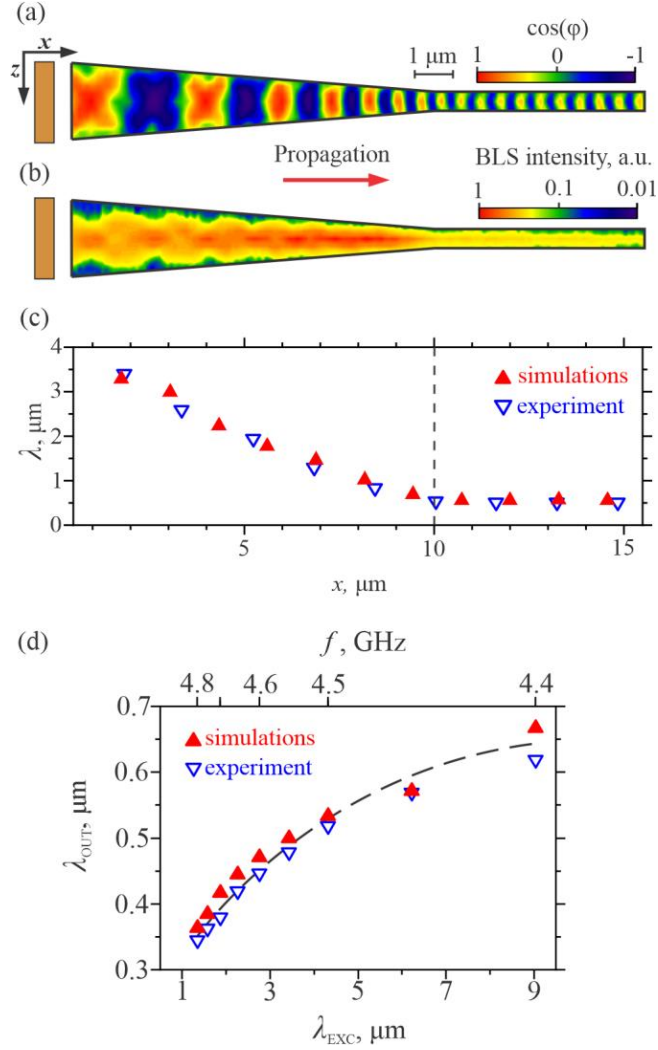

FIG. 2. Representative maps of the spin-wave phase (a) and intensity (b) recorded by BLS at the excitation frequency $f$=4.5 GHz. The left edge of the maps corresponds to the displacement $x$=0.5 μm from the edge of the antenna. (c) Spatial dependence of the wavelength of the spin wave with the frequency $f$=4.5 GHz. Vertical dashed line marks the end of the transition region. (d) Spin-wave wavelength at the end of the transition region, $\lambda_{OUT}$, as a function of the wavelength of the spin wave excited by the antenna, $\lambda_{EXC}$. Dashed curve – guide for the eye. In (c) and (d): point-down triangles – experimental data, point-up triangles – results of micromagnetic simulations. The data are obtained at $H_0$=1000 Oe.

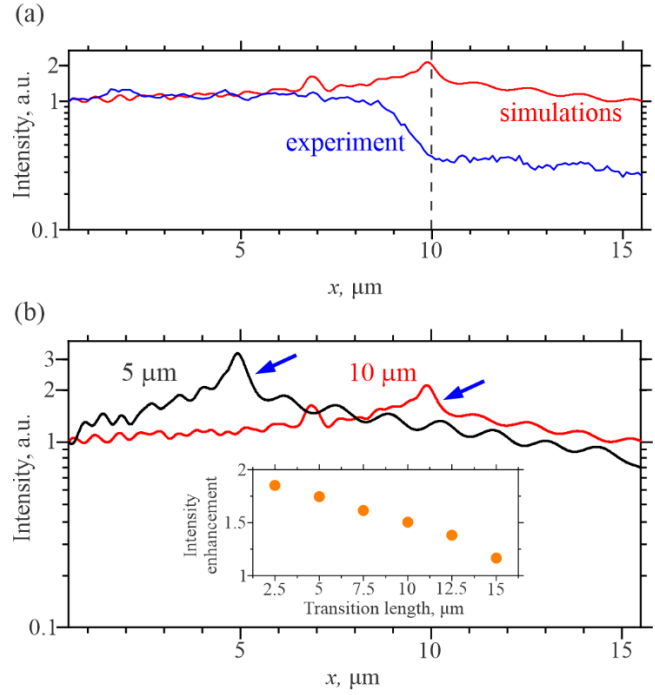

FIG. 3. (a) Spatial dependences of the spin-wave intensity integrated over the width of the waveguide obtained from the measurements and micromagnetic simulations, as labelled. Vertical dashed line marks the end of the transition region. (b) Spatial dependence of the spin-wave intensity calculated for the waveguides with the transition length of 10 and 5 μm, as labelled. Arrows mark the intensity drop due to reflections at the end of the transition region. Inset shows the ratio between the intensity detected at a point 0.5 μm beyond the end of the transition region and the intensity detected at its beginning as a function of the transition length. The data are obtained at $H_0$=1000 Oe and $f$=4.5 GHz.

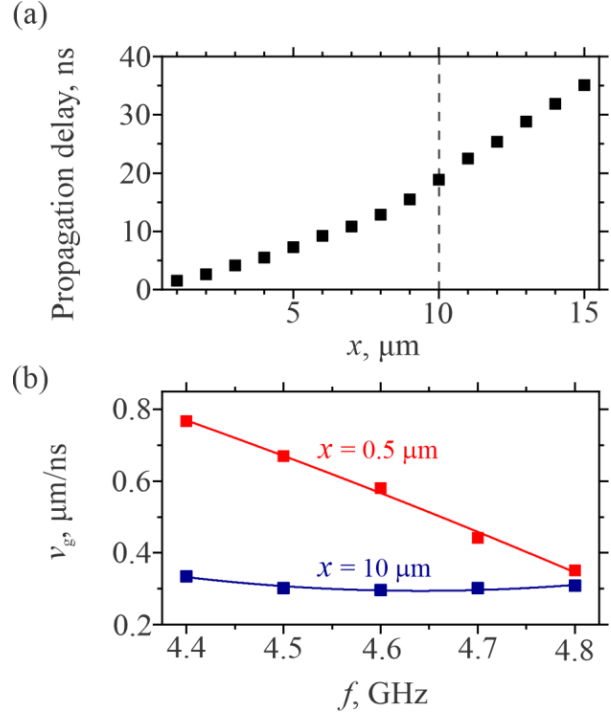

FIG. 4. Spatial dependence of the propagation delay measured for a 20-ns long spin-wave pulse at the carrier frequency $f$=4.5 GHz. Vertical dashed line marks the end of the transition region. Frequency dependence of the group velocity at the beginning of the transition region ($x$=0.5 μm) and at its end ($x$=10 μm), as labelled. Symbols – experimental data. Curves – guide for the eye. The data are obtained at $H_0$=1000 Oe.